\documentclass[12pt, a4paper]{article}
\usepackage{graphicx}
\usepackage{latexsym}
\usepackage{amsmath}
\usepackage{amssymb}
\usepackage{epsfig}

\begin{document}

\title{Josephson junctions and dark energy}

\author{Philippe Jetzer and Norbert Straumann\\
        Institute for Theoretical Physics University of Zurich,\\
        Winterthurerstrasse 190, \\
        CH--8057 Zurich, Switzerland}

\maketitle

\begin{abstract}
In a recent paper Beck and Mackey [astro-ph/0603397] argue that
the argument we gave in our paper [Phys. Lett. B {\bf 606}, 77
(2005)] to disprove their claim that dark energy can be discovered
in the Lab through
noise measurements of Josephson junctions is incorrect. 
In particular, they emphasize that the measured noise spectrum
in Josephson junctions is a consequence of the fluctuation
dissipation theorem, while our argument was based on equilibrium
statistical mechanics.
In this note we show that the fluctuation dissipation relation
does not depend upon any shift of vacuum (zero-point) energies,
and therefore, as already concluded in our previous paper, dark energy
has nothing to do with the proposed measurements.

\end{abstract}

\section{Introduction}

It has been suggested by Beck and Mackey \cite{BM1} that part of the
zero-point energy of the radiation field that is gravitationally
active can be determined from noise measurements of Josephson
junctions. This caused some widespread attention.  In a reaction we
\cite{JS} we thought we had clearly shown that there is no basis for
this claim, by following the reasoning in \cite{BM1} for a much
simpler model, for which it is very obvious that the authors
misinterpreted their formulae. More  generally, we stated that the
absolute value of the zero-point energy of a quantum mechanical
system has no physical meaning as long as gravitational coupling is
ignored. All that is measurable are \emph{changes} of the zero-point
energy under variations of system parameters or of external
couplings, like an applied voltage.

Recently, Beck and Mackey argued \cite{BM2} that our argument does
not apply to their original proposal. They state in particular that
the measured noise spectrum of Josephson junctions is a consequence
of the fluctuation-dissipation theorem, while our argument `` is
based on equilibrium statistical mechanics and does not incorporate
non-equilibrium effects''.

Mainly for this reason we demonstrate below explicitly that the
fluctuation-dissipation relation is immune to any shift of vacuum
(zero-point) energies. In particular, the vacuum energy of the
radiation field, cut-off at some frequency, does not enter in the
fluctuation-dissipation theorem. The misinterpretation of the
formulae by Beck and Mackey is, as we shall show, exactly of the
same type we pointed out in our first reply \cite{JS}.

\section{On the fluctuation-dissipation theorem}

For the sake of our argument, we have to briefly recall a derivation
of the famous \emph{fluctuation-dissipation theorem}, originally
discovered by Callen and Welton \cite{CW}.

Mathematically, the fluctuation-dissipation theorem relates for an
operator $Q(t)=e^{iHt}Q e^{-iHt}$ ($H$ the Hamiltonian of the
system) the canonical expectation values (denoted by
$\langle\cdot\rangle$) of the commutator $[Q(t),Q(0)]_-$ and the
anti-commutator $[Q(t),Q(0)]_+$.

Let $\alpha^{''}(\omega)$ be the Fourier transform of
$\frac{1}{2}\langle [Q(t),Q(0)]_-\rangle$,
\begin{equation}
\alpha''(\omega)=\frac{1}{2}\int_{-\infty}^\infty \langle
[Q(t),Q(0)]_-\rangle e^{i\omega t}~dt.
\end{equation}
In physical applications $\alpha''(\omega)$ is the imaginary part of
a generalized susceptibility $\alpha(\omega)$ (Kubo formula for
linear response), and describes energy dissipation caused by an
external perturbation. It is now a matter of at most a few
lines\footnote{All that one needs is that the Fourier transform of
$\langle Q Q(t)\rangle$ is equal to that of $\langle Q(t)Q\rangle$ times
$e^{-\beta\omega},~\beta=1/kT,~\hbar=1$.} to show that
\begin{equation}
\frac{1}{2}\langle [Q(t),Q(0)]_+\rangle=\frac{1}{\pi}\int_0^\infty
\coth\frac{\beta\omega}{2}\alpha''(\omega)e^{-i\omega t}~d\omega
\end{equation}
($\beta=1/kT,~\hbar=1$).

For $t=0$ this formula expresses the fluctuation $\langle
Q^2\rangle$ in terms of the susceptibility. A slightly different
form is
\begin{equation}
\langle Q^2\rangle=\frac{2}{\pi}\int_0^\infty
\left[\frac{1}{2}\omega+\frac{\omega}{e^{\beta\omega}-1}\right]\alpha''(\omega)\frac{d\omega}{\omega}.
\end{equation}
This connection is what one calls the fluctuation-dissipation
theorem. The formula used by Beck and Mackey is equivalent to (3).
They use the relation between $\langle\dot{Q}^2\rangle$ and
$\alpha''(\omega)$, which is obtained from (3) by multiplying the
integrand with $\omega^2$.

Obviously, $\langle [Q(t),Q(0)]_{\pm}\rangle$ do not change under
the substitution  $H \rightarrow H+const.$ Therefore, this
substitution does not induce an additive constant in the square
bracket of (3), contrary to what was suggested by Beck and Mackey in
their equation (5.13). The term $\frac{1}{2}\omega$ in the square
bracket has nothing to do with the ground state energy of the
system. The latter does not show up in the fluctuation-dissipation
theorem, and can be treated as an arbitrary normalization.

Testing relations like (3) amounts to testing basic quantum theory,
and has nothing to do with dark energy.

\section{Concluding remarks}

We hope to have shown in sufficient detail that \emph{vacuum
(zero-point) energies do not show up in any application of the
fluctuation-dissipation theorem}. Therefore, the change suggested in
eq. (5.13) of Ref. \cite{BM2}, is wrong. The term
$\frac{1}{2}\omega$ in the square bracket of (3) has nothing to do
with vacuum energies.

In their rebuttal \cite{BM2}, Beck and Mackey make in Sect.~V the
following general statement: `` The theory of dissipative
non-equilibrium quantum systems, such as driven Josephson junctions,
is much less well understood than the Casimir effect. Whether the
dissipative quantum theory underlying resistivity shunted Josephson
junctions can be renormalized is presently unclear. Hence the
absolute value of the vacuum energy may well have physical meaning
for these kinds of superconducting quantum systems.'' Our comment to
this -- beside our concrete remarks in the previous section -- is:
The basic theory underlying resistivity shunted Josephson is a
renormalizable theory, namely quantum electrodynamics. In this
theory the vacuum energy can be normalized to any value, without
changing any observable prediction. This can only change when an
enlarged theory containing gravity is considered. Outside gravity,
vacuum energies are unmeasurable. (The standard model of particle
physics is renormalizable.)

\end{document}